\documentclass[%
 aip,
rsi,%
 amsmath,amssymb,
 reprint,%
]{revtex4-1}
\usepackage{siunitx}
\usepackage{graphicx}
\usepackage{dcolumn}
\usepackage{bm}
\begin{document}

\preprint{AIP/123-QED}

\title{Laser Compression via fast-extending plasma gratings}
\author{Zhaohui Wu}
\author{Xiaoming Zeng}
\author{Zhaoli Li}
\author{Zhimeng Zhang}
\author{Xiaodong Wang}
\author{Bilong Hu}
\author{Xiao Wang}%
\author{Jie Mu}
\author{Jingqin Su}
\author{Xiaofeng Wei}
\author{Yanlei Zuo}%
\email{zuoyanlei@tsinghua.org.cn}
\affiliation{Science and Technology on Plasma Physics Laboratory, Research Center of Laser Fusion, China Academic of Engineering Physics, Mianyang, Sichuan, China, 621900.}


\date{\today}

\begin{abstract}
  It is proposed a new method of compressing laser pulse by fast extending plasma gratings(FEPG), which is created by ionizing the hypersound wave generated by stimulated Brillouin scattering(SBS) in the background gas. Ionized by a short laser pulse, the phonon forms a light-velocity FEPG to fully reflect a resonant pump laser. As the reflecting surface moves with a light velocity, the reflected pulse is temporally overlapped and compressed. This regime is supported by the simulation results of a fully kinetic particle-in-cell(PIC) code Opic with a laser wavelength of 1$\mu$m, displaying a pump pulse is compressed from 13ps to a few cycles(7.2fs), with an efficiency close to 80\%. It is a promising method to produce critical laser powers due to several features: high efficiency without a linear stage, robustness to plasma instabilities, no seed and a wide range of pump intensity.
\end{abstract}

\maketitle

Compression of ultrashort laser pulses by plasma has the potential to boost the instantaneous laser power to new levels due to the robustness of plasma in strong laser fields. Currently, stimulated raman backscattering(SRBS) amplification \cite{Malkin991,Malkin00,Trines111,Ping04,Cheng05,JUN07,Vieux17,Wu20} and strongly coupled stimulated Brillouin scattering (SCSBS)\cite{Andreev06,Lancia13,Lancia16,Peng16,Marques19} are two promising methods for this application. In both SRBS and SCSBS, the energy of relative longer pump pulses are scattered to conterprorogating short seeds pulse by the plasma waves, Langmuir wave or ion-acoustic wave, respectively. Theoretical results show the possibility for a final amplified seed peak intensity of over $10^{16}~\rm{W/cm^2}$, which is about 5 orders of magnitude larger than the maximum intensity obtained by a solid-state crystal.

However, both BRA and scSBS have to experience a linear stage, in which the plasma waves grow up from density perturbation. In this stage, the pump is little depleted and the seed pulse is stretched, so both the transfer efficiency and compression ratio are degraded. Moreover, due to the influence of plasma instabilities, the linear stage in the experiment was usually much longer than expectation. Therefore, the benefits of the theoretical prediction are still difficult to be realized. Although a strong seed helps to achieve the nonlinear stage quickly, it has to pay the price of reducing the compression ratio.

\begin{figure}[htpb]
\centering
\includegraphics[width=3 in]{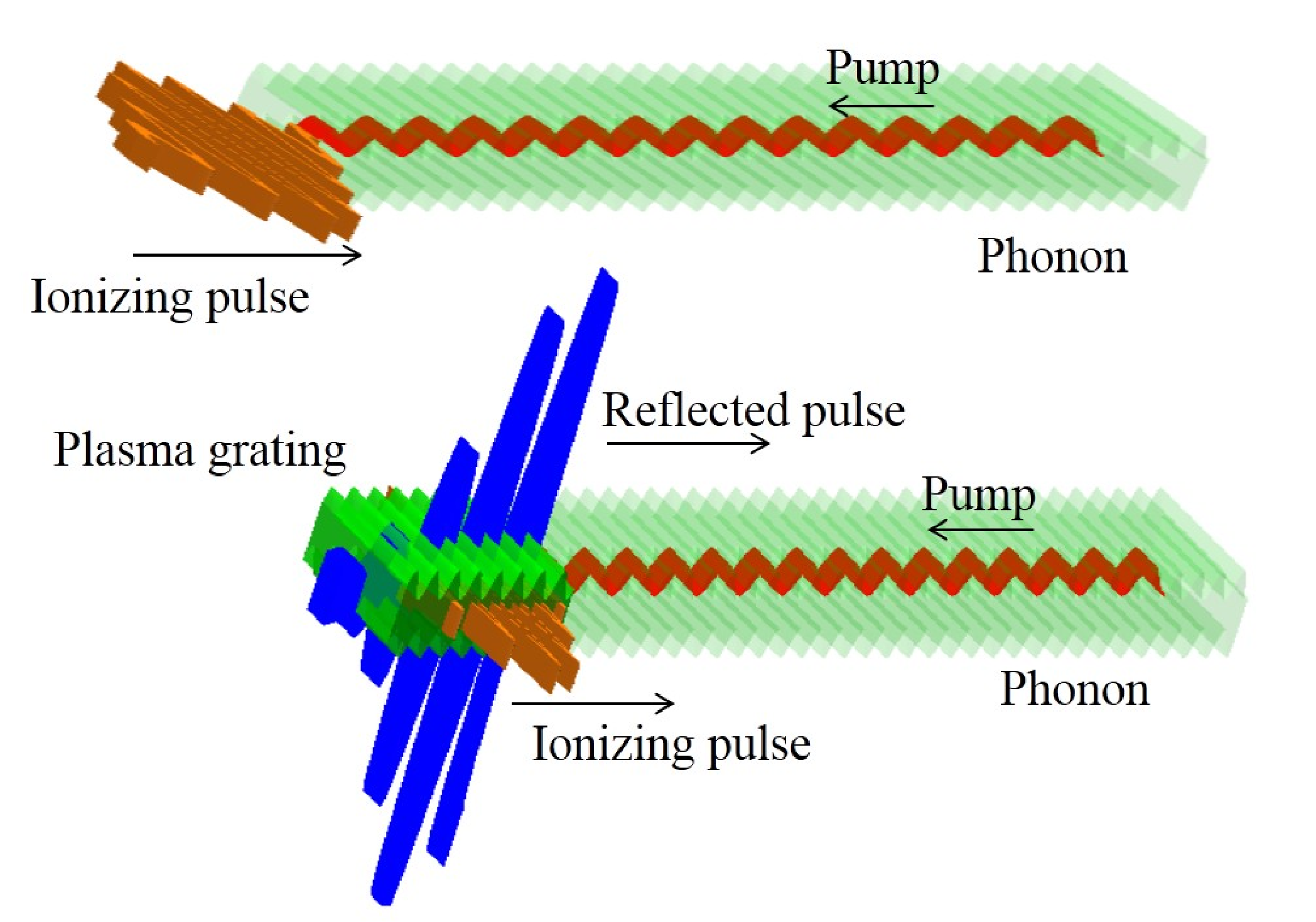}
\caption{Illustration of the laser compression by fast-extending plasma grating produced by a  short pulse(brown pulse, cross-polarized) and the phonon(transparent green fringes) in background gas. As boundary of the plasma grating(green fringes) moves with the ionizing pulse, the reflected pulse(blue pulse) of the pump(red pulse) is compressed.
\label{fig:scheme}}
\end{figure}

 The difficulties could be overcame by using a novel way, named fast-extending plasma-grating compression(FEPGC). The scheme is shown in Fig\ref{fig:scheme}. The phonon---a hypersound wave with a spatial period half wavelength of the stimulating laser---is supposed to be created by the stimulated Brillouin scattering(SBS) of 2 counter-prorogating laser pulses in gas, which has bean widely found in $CH_4$, $N_2$ and $H_2$\cite{Hpop}. A pump pulse is set below the ionization threshold of the background gas, so it can pass through the phonon in which variation of the refraction index is still small. By using a short pulse to ionize the phonon, a plasma grating extending with light-velocity is created. As the refraction index is much more sensitive to density in plasma than gas, the pump pulse satisfying the Brag condition of the grating would be fully reflected back. To be noted is that the reflecting surface moves with a light-velocity, so the reflected pulse(RP) is temporally overlapped to a short one. As the plasma grating is produced by ionizing the phonon but not grows from density perturbations, the pump can be transiently depleted and thereby the compression process has not a linear stage.

Plasma gratings created by laser-plasma interaction have been widely studied for manipulating high-power lasers\cite{Forslund75,Chunhui05,Chunhui06,Lehmann16,Lehmann17,Lehmann18,Peng19,Peng191}. However, they are usually difficult to be controlled due to plasma instabilities. Take a different approach, we consider generating plasma gratings from more-stable acoustic waves. The hypersound wave excited and maintained by the SBS effect in gas can be described by a material equation\cite{SBS}
\begin{eqnarray}
\begin{array}{rcl}
\frac{\partial \rho}{\partial t}+(-i\delta\omega+\frac{\Gamma_B}{2})\rho=-\frac{i\gamma_e\varepsilon_0k_B}{4\upsilon}E_0E_1^*\\
\end{array}
\label{acousticwave}
\end{eqnarray}

where $E_{0,1}$ and $\omega_{0,1}$ are the E-field and frequencies of the laser pulses;
$n$ is the refraction index of the gas; $c$ is the light velocity in vacuum; $\varepsilon_0$ is the permittivity of free space;
$\gamma_e=\rho_0(\partial\varepsilon/\rho)$ is the electrostriction coefficient, and it can be written as $(n^2-1)(n^2+2)/3$ in centrosymmetric materials;  $k_B$ is the Brillouin
wavevector of the acoustic wave; $\rho$ is the amplitude of the density perturbation; $\rho_0$ is the medium density; $\Gamma_B$ is the inverse of the acoustic lifetime $\tau_B$ given by

\begin{eqnarray}
\begin{array}{rcl}
\Gamma_B=\frac{1}{\tau_B}=\frac{k_B^2\eta}{\rho_0}
\end{array}
\label{gammB}
\end{eqnarray}
where $\eta$ is the is the viscosity. Assuming the $E_0$, $E_1$ and density wave  satisfy the resonant condition, $\rho$ has a steady-state value for a sufficient interaction time
\begin{eqnarray}
\begin{array}{rcl}
\frac{\rho}{\rho_0}=\frac{I\gamma_e}{\upsilon cnk_B\eta}
\end{array}
\label{steadyrho}
\end{eqnarray}
where, $I=\varepsilon_0 c n E_0 E^*_1/2$ is the interfered intensity of the laser pulses, and $\upsilon$ is the acoustic velocity. By employing the parameters of several gases, the steady-state $\rho/\rho_0$ are shown in Tab.\ref{tab:densitywave}. A particular value of $\rho/\rho_0$ can be obtained by adjusting the laser intensity  which is mainly limited by the threshold of multiphoton ionization. For hydrogen, the measurable ionization was observed at an intensity of some $10^{13}~\rm{W/cm^2}$ with laser wavelength of 1064 nm\cite{Wolff88}.

\begin{table}[htpb]
\centering
\caption{\label{tab:densitywave} Calculation of steady-state $\rho/\rho_0$ for several gases. Both the laser pulses have the wavelength of 1 $\mu$m. $I$ is used the unit of $\rm{10^{12}~W/cm^2}$. The gas pressure is $P=1~atm$ temperature is $T=20^\circ C$.}
\begin{tabular}{cccccc}
\hline
Gas & $\eta(10^{-6}\rm{Pa.s})$ & $\upsilon(\rm{m/s})$ &n &$\gamma_e(10^{-4})$ & $\rho/\rho_0$ \\
\hline
\itshape $H_2$ &8.8& 1295&1.00013 & 2.6 &0.06$I$\\
\itshape $CH_4$ &10.8& 440& 1.00044 & 8.8 & 0.47$I$\\
\itshape $N_2$ &$16.6$&334 & 1.00030 & 6.0 & 0.28$I$\\
\hline
\end{tabular}
\end{table}

After the phonon is ionized, the laser-plasma interaction can be depicted by three-waves equations as follows\cite{Kruer88}:

\begin{eqnarray}
\begin{array}{rcl}
(\frac{\partial^2}{\partial t^2}-c^2\Delta^2+\omega^2_{pe})\textbf{A}_a=-\frac{4\pi e^2}{m_e}\delta n_e\textbf{A}_b\\
(\frac{\partial^2}{\partial t^2}-c^2\Delta^2+\omega^2_{pe})\textbf{A}_b=-\frac{4\pi e^2}{m_e}\delta n_e\textbf{A}_a\\
(\frac{\partial^2}{\partial t^2}-c^2_s\Delta^2)\delta n_e=\frac{Ze^2{n_e}_0}{m_em_ic^2}\Delta^2(\textbf{A}_a\textbf{A}_b)
\end{array}
\label{threewave0}
\end{eqnarray}

where $\textbf{A}_{a,b}$  are laser vector potentials; $c_s=\sqrt{ZT_e/m_i}$ is the ion acoustic velocity; $Z$ is the plasma ion charge; $T_e$ is the plasma temperature; $e$ is the electron charge; $m_{e(i)}$ is the electron(ion) mass; $\delta n_e$ is the amplitude of electron density perturbation, and ${n_e}_{0}$ is the average electron density.

By using the slowly-varying envelope approximation, the 1-dimensional Eq.\ref{threewave0} can be simplified to
\begin{eqnarray}
\begin{array}{rcl}
&&a_t+2a_{\zeta} =-Vb f\\
&&b_t=Va f^*\\
&&(\partial_t+{\partial_\zeta}-i\frac{k^2_Bc^2_s-\omega_B^2}{2\omega_B})f=\frac{Ze^2}{m_em_ic^2}ab^*
\end{array}
\label{threewave1}
\end{eqnarray}
in variable $\zeta=z/c+t$, where $a$ and $b$ are the normalized amplitude of the pump and RP, $V\equiv i\frac{\omega^2_{pe}}{2\omega_a}$ is coupling constant; $\omega_{pe}=\sqrt{n_ee^2/m_e\varepsilon_0}$ is plasma frequency; $f\equiv\frac{\delta n_e}{{n_e}_{0}}$ is the normalized plasma density. As the plasma grating is directly obtained by ionizing the phonon, the acoustic frequency  $\omega_B=v/\lambda$ and wave number  $k_B$ are remained. The RP would still satisfy the resonant condition $\omega_b=\omega_a-\omega_B\approx\omega_a$ and $k_B=k_a+k_b\approx2k_a$, where $\omega_{a,b}$ and $k_{a,b}$ are the laser frequencies and wave numbers. Commonly, it has $v\ll c_s$, so the growth of $f$ has a frequency detuning of $(k^2_Bc^2_s-\omega_B^2)/{2\omega_B}$ , as shown in Eq.\ref{threewave1}. As the boundary of the plasma grating moves with a light velocity, the boundary condition can be written as $f(\zeta=\zeta_0)=f_0=Z\rho/\rho_0$, where $\zeta_0$ is the position where the gas starts to be fully ionized. The laser-plasma interaction in the partly-ionized section is neglected because it can be very short by using a sharp-front ionizing pulse. Moreover, for the plasma grating, the laser reflection becomes significant only when the average density is close to the Brag condition.

The compression process can be described by numerically solving the Eq.\ref{threewave1}. However, in the limit of a low pump intensity below the ionization threshold, the material equation in Eq.\ref{threewave1} can be neglected for several reasons. Firstly, the resonant laser exponentially decaying in the plasma grating can only enter into a very thin plasma layer. Secondly, as the ionization pulse moves forward, the plasma layer is continually replaced by new ones, resulting in a very short interaction time for laser and plasma interaction, which is down to a few optical cycles. Due to the participation of heavy ion, the density variation driven by laser is quite small compared with the plasma grating.

As the interaction directly enters into the nonlinear stage, it has $\frac{\partial a}{\partial t}\ll \frac{\partial a}{\partial \zeta}$, so Eq.\ref{threewave1} is given as
\begin{eqnarray}
\begin{array}{rcl}
&&2a_{\zeta} =-Vb\delta f\\
&&b_t=Va\delta f^*\\
&&f=f_0
\end{array}
\label{threewave2}
\end{eqnarray}

Eq.\ref{threewave2} can be written as
\begin{eqnarray}
\begin{array}{rcl}
a_{\zeta t}=V^2 f_0^2a/2
\end{array}
\label{threewave3}
\end{eqnarray}

It can be approximated by a self-similar function $a(\zeta,t)=\tilde{a}(\sqrt{\xi})$, $\xi\equiv Vf_0\sqrt{2\zeta t}$, that satisfies            the zero-class Bessel function
\begin{eqnarray}
\begin{array}{rcl}
\tilde{a}_{\xi\xi}+\tilde{a}_{\xi}/\xi=\tilde{a}
\end{array}
\label{threewave4}
\end{eqnarray}

With the boundary condition of $a(\xi=0)=a_0$, the pump has a solution $a=a_0J_0(\xi)$, then
\begin{eqnarray}
\begin{array}{rcl}
b=2a_0Vf_0J_1(\xi)t/\xi.
\end{array}
\label{SBS}
\end{eqnarray}

Let $b_t=Va\delta f^*=0$, the RP has the maximum value at the first peak for $\xi_M\approx2.4$, gives
\begin{eqnarray}
\begin{array}{rcl}
b_{M}\approx0.43 a_0Vf_0t.
\end{array}
\label{SBSM}
\end{eqnarray}

In the ideal case, $b_M$ grows linearly with time. However, just as BRA and sc-SBS, the output intensities are limited by several plasma instabilities. The modulation instability corresponds the variations in plasma density driven via the ponderomotive force by intensity modulations in the light beam, with a growth rate of $\sim b{^2}\omega{^2_{pe}}/2\omega_a$\cite{Malkin991}. The longitudinal modulation instability can make the RP peak depart from the maximum gain and gradually tend to saturation, while the transversal modulation instability suppresses the compression mainly by causing filamentation. Combining with Eq.\ref{SBSM}, it gives the limited time of $t_{md}=\frac{2.53\Lambda_{md}}{V}(a_0f)^{-2/3}$, where $\Lambda_{md}$ is the number of exponentiations which can be tolerated for modulation. Furthermore, as the laser pulse is compressed to a few optical cycles, the group velocity dispersion(GVD) given as $\beta=\frac{\partial^2\omega \bar{n}}{c\partial\omega}|_{\omega=\omega_0}\approx\frac{\omega^2_{pe}}{\omega^3_0c}$ can not be neglected. For a transform limit pulse with $T_0$, the pulse is stretched to $T=\sqrt{T^2_0+\beta ct/T^2_0}$. It limits the shortest pulse can be obtained by the compression.

The stimulated Raman backscattering(SRBS) from the pump and forward Raman scattering(FRS) from the RP have the same interaction time as RP. Both of them are based on Langmuir wave, with growth rates of $\gamma_b=a\sqrt{\omega_{pe}\omega_a}/2$ and $\gamma_{fw}=b\sqrt{\omega^3_{pe}/8\omega_a}$\cite{Forslund75,Kruer88}, respectively. The growth of Langmuir waves are strongly damped by the frequency detuning due to the strong density inhomogeneity of the plasma grating. The effective growth rates of these 2 instabilities can be roughly estimated as $\gamma^2/\Delta\omega_{pe}$\cite{Malkin14}, where $\Delta\omega_{pe}$ is the frequency variation of Langmuir wave due to plasma inhomogeneity.

Several major plasma instabilities suppressing the plasma gratings can be avoided here. Plasma kinetic effected is significantly migrated due to a weak plasma-heating effect. In the scheme of FEPGC, the pump is set below the ionization of background gas. The compression is just behind the ionizing process, leaving a very short distance before the RP for plasma heating and precursors generation by the pump. The regime is quite similar with seed-ionizing\cite{Malkin001,clark02,zhang17} or flying-focus\cite{Turbunll18,Dustin18,Sainte} BRA. Moreover, the collision and Landau damping have little threat to the compression because they can hardly impact the plasma grating within the interaction time of the laser and plasma layer(a few optical cycles). In addition, the wavebreaking is far from approach. The wavebreaking has a time of $t_{wb}=\sqrt{2m_i/m_e}/\omega_ab\approx60/\omega_ab$ for hydrogen\cite{Forslund75,Huller91}, showing the toleration of laser amplitude can be up to $b>1$ for a duration of a few optical cycles.

To be noted is that the Brag condition of the grating is varied after the ionization. In a typical Brag grating, the Brag condition is written as $\Lambda=m\lambda_0/2\bar{n}$, where $\Lambda$ is the spatial period of the grating, $m$ is the diffractive number, $\lambda_0$ is the resonant laser wavelength in gas, $\bar{n}$ the average refraction index. $\bar{n}\approx1$ in the background gas, so $\Lambda\approx\lambda_0/2$ . After the gas is ionized, the refraction index in plasma becomes $\bar{n}=\sqrt{1-\frac{{n^2_e}_{0}e^2}{m_e\varepsilon_0\omega^2}}<1$, indicating the resonant laser wavelength is slightly down shifted. That's why the reflectivity decreases after the gas is ionized in the SBS experiment\cite{Gorbunov83}. The pump with a small incidence angle or a shorter wavelength length can satisfy the new Brag condition. For instance, in hydrogen with $\rho_0=0.01n_c$, where $n_c=\omega^2_am_e\varepsilon_0/e^2$, it requires $\theta=8.1^{\circ}$ or $\lambda_a=0.99\lambda_0$, where $\theta$ is the incidence angle and $\lambda_a$ is the resonant wavelengths in plasma, respectively.

By employing hydrogen with a hypersound wave as the background gas, the ionization and compression process are simulated with a fully kinetic particle-in-cell code Opic\cite{zhang12,Zhang14,zhang17}. The simulation was carried out in a moving window of $800\lambda_a$ with cells of $\Delta_z=\lambda_a/20$, and $32$ particles for each cell. A 10$T$ 800-nm short pulse with a peak amplitude of 0.03 is used for ionization pulse, where $T$ is one optical cycle of $\lambda_a$. It has a crossed polarization with the pump pulse in order not to affect the compression. The ionization process is represented by an Ammosov-Delone-Krainov tunnel ionization model(ADK). The pump pulse has a wavelength of $\lambda_a=1~\mu m$ and an amplitude below the ionization threshold. To satisfy the Brag condition for the pump, an optimal $\Lambda>0.5\lambda_a$ of the density wave is obtained according to the maximum pump depletion in the simulation.

\begin{figure}[htpb]
\centering
\includegraphics[width=1.6 in]{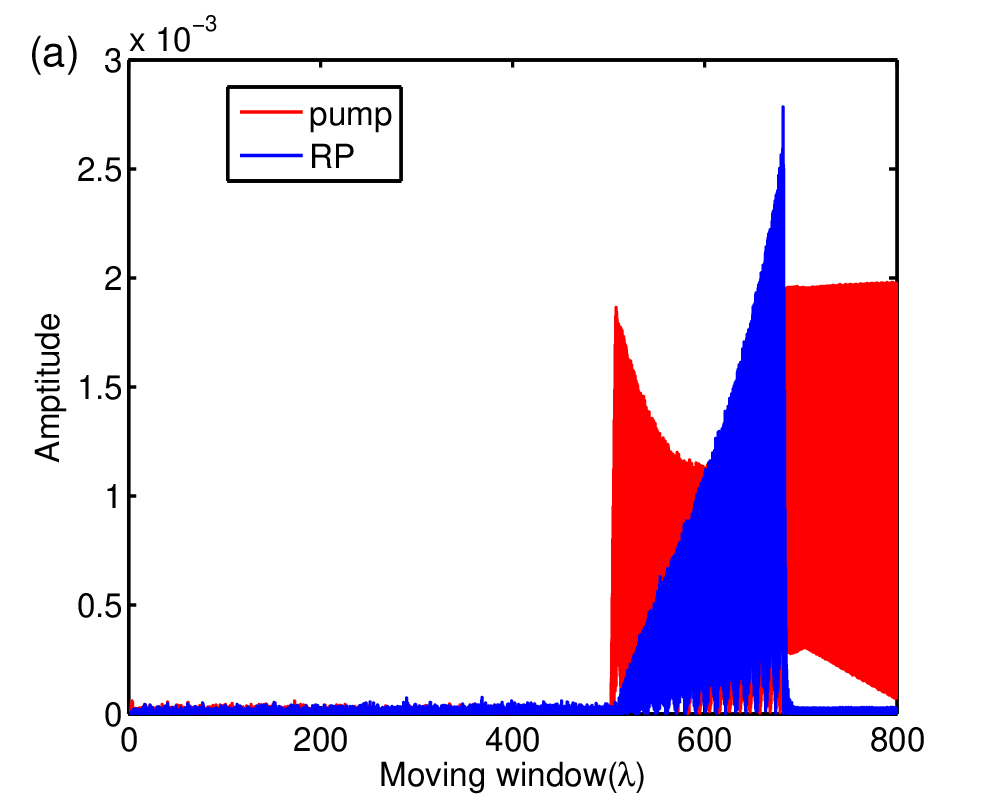}
\includegraphics[width=1.6 in]{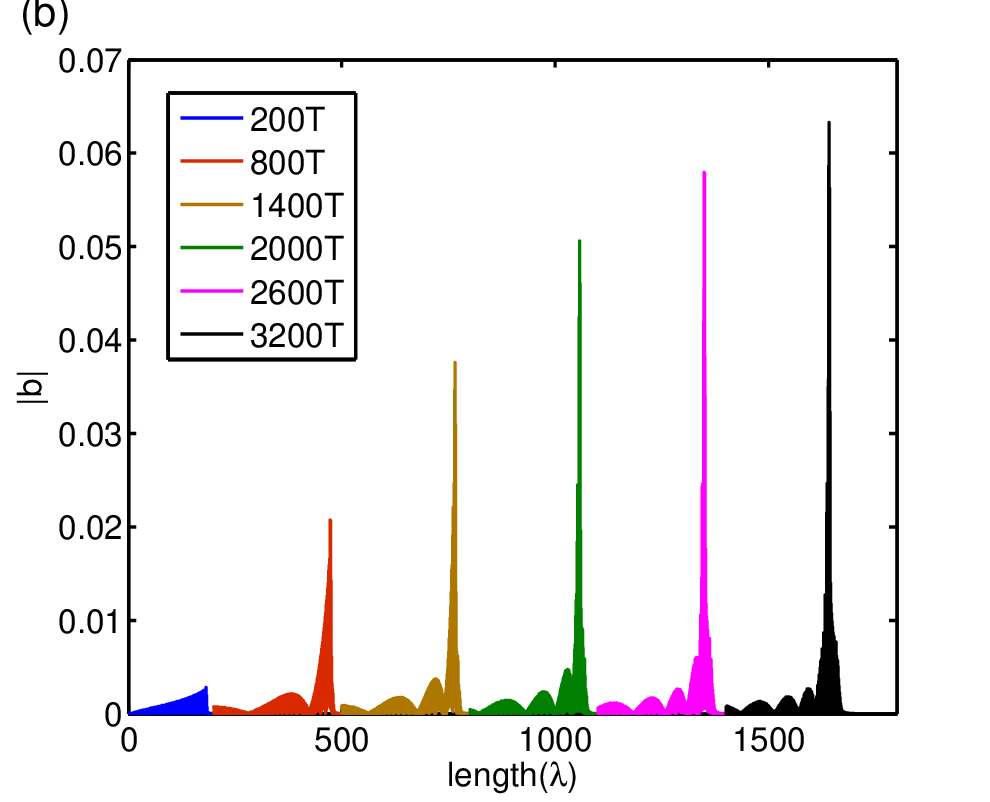}\\
\includegraphics[width=1.6 in]{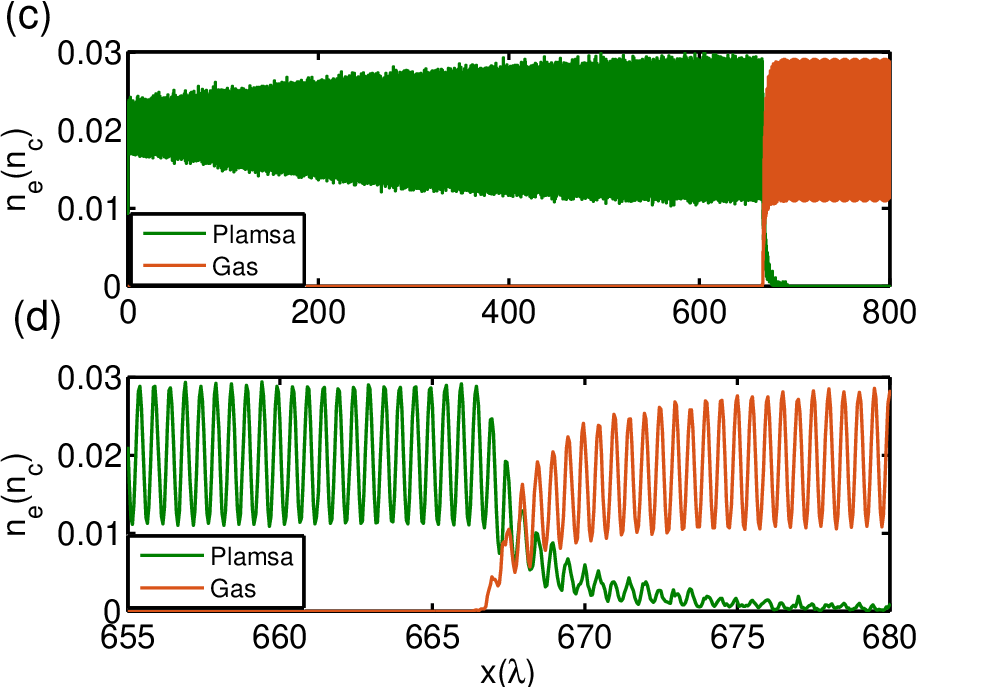}
\includegraphics[width=1.6 in]{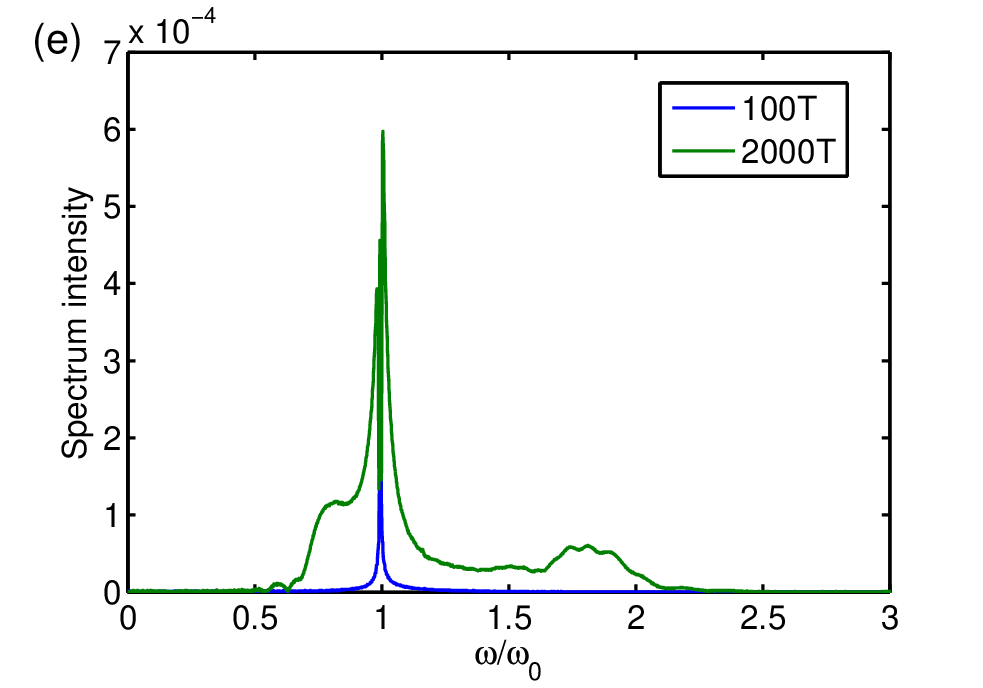}\\
\caption {PIC Simulation results of pulse compression by FEPG at $\rho_0=0.01n_c$, $\rho/\rho_0=0.5$, $a_0=0.002$, $\Lambda=0.50238\lambda_a$, and $\lambda_a=1~\mu m$, corresponding to $T=3.3~fs$. (a) The pump(red) and RP(blue) at t=100~$T$. (b) Evolution of the RP shape at different interaction time. Overall(c) and specific(d) distribution of plasma grating and background gas at t=2000~$T$. (e) The spectrum of SBS at $t=100~T$ and $t=2000~T$.
\label{fig:mainsimulation}}
\end{figure}

The simulation results are displayed in Fig\ref{fig:mainsimulation}. The pump pulse is swiftly depleted soon after it encounters the plasma grating, implying the nonlinear stage can be directly reached, as shown in Fig\ref{fig:mainsimulation}(a). The initial RP at $t=200T$ has a half-maximum full width(HMFW) duration around 80 fs, and then is further compressed, with an amplitude far exceeding that of the pump. The shortest duration of 7.2 fs is obtained at the interaction time of $2000T$, which is close to 2 cycles of the pump wavelength. After that, the duration is slightly returned back to around 9 fs, but the amplitude still grows up to 0.07 until $t=3200T$, with a compression efficiency close to 80\%. The overall and specific profiles of plasma grating are shown in Fig\ref{fig:mainsimulation}(c)-(d). It has a very sharp boundary between the background gas and plasma zone. The density fluctuation gradually decreases as it is far from the boundary, because of the thermodynamic motion of the plasma. However, the compression is not impacted because only a very thin plasma layer in Fig.\ref{fig:mainsimulation}(d) participates the interaction. The RP spectrum displays an extremely wide range from $0.5\omega_a-2\omega_a$ at $t=2000~T$, implying a short pulse of a few cycles can be supported, as shown in Fig\ref{fig:mainsimulation}(d).

\begin{figure}[htpb]
\centering
\includegraphics[width=1.6 in]{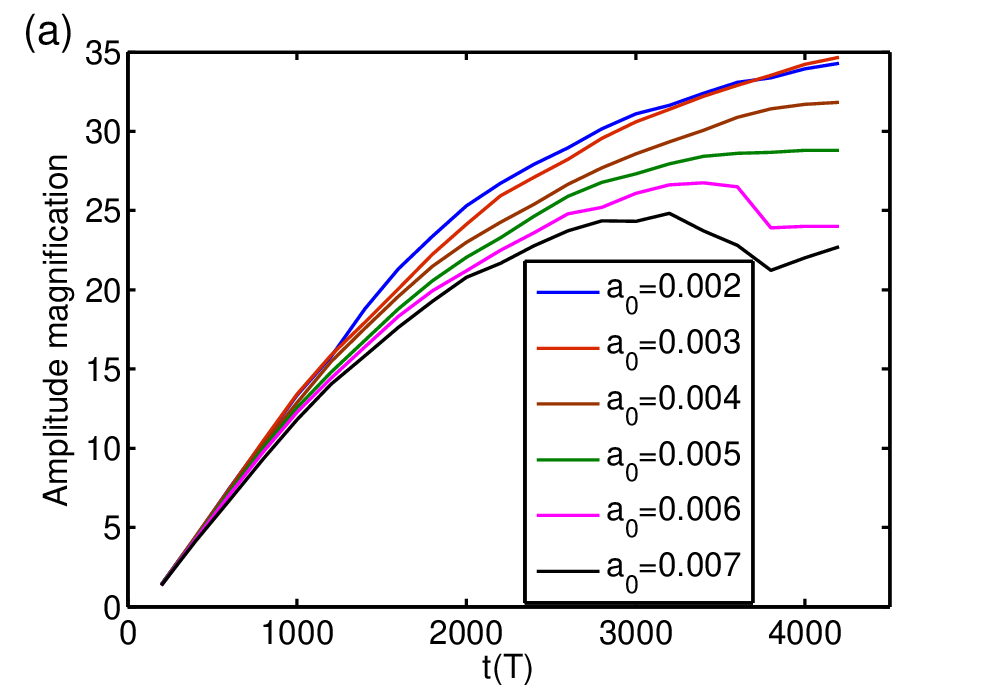}
\includegraphics[width=1.6 in]{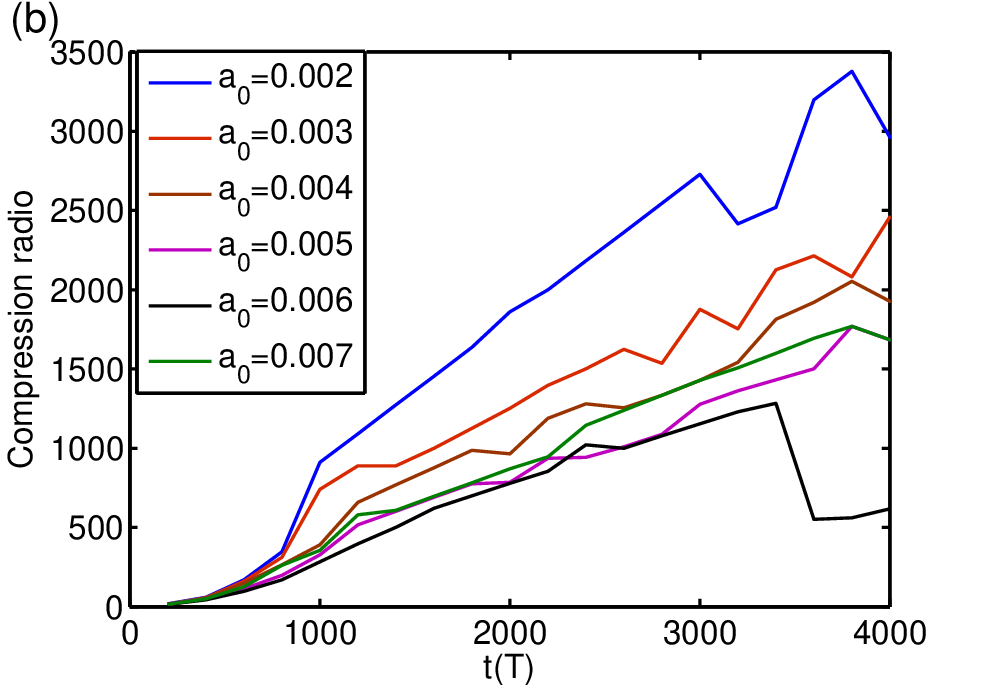}\\
\includegraphics[width=1.6 in]{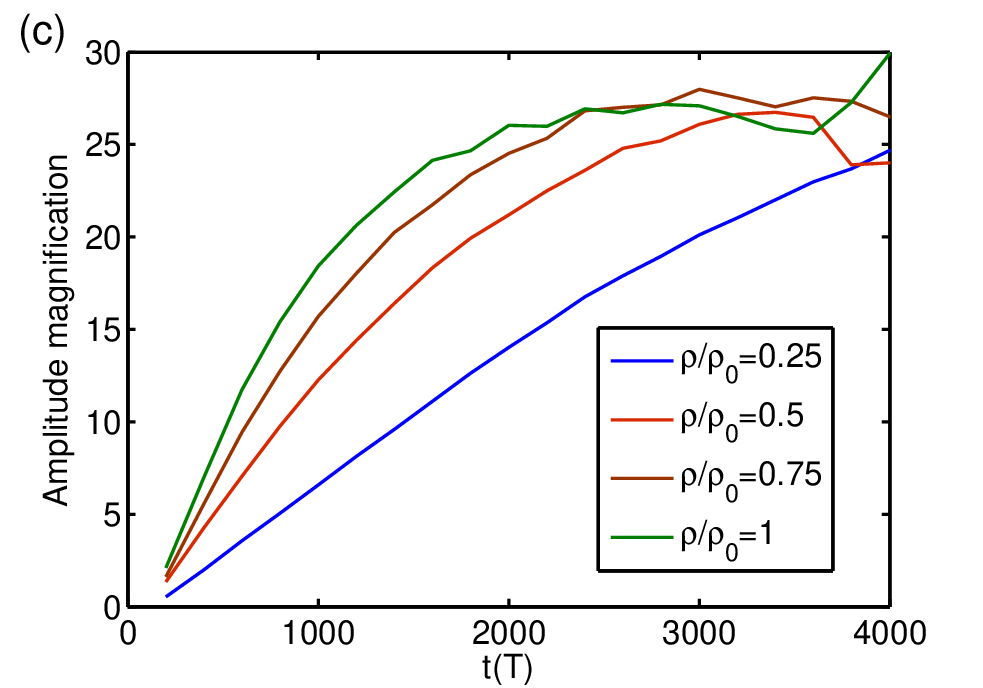}
\includegraphics[width=1.6 in]{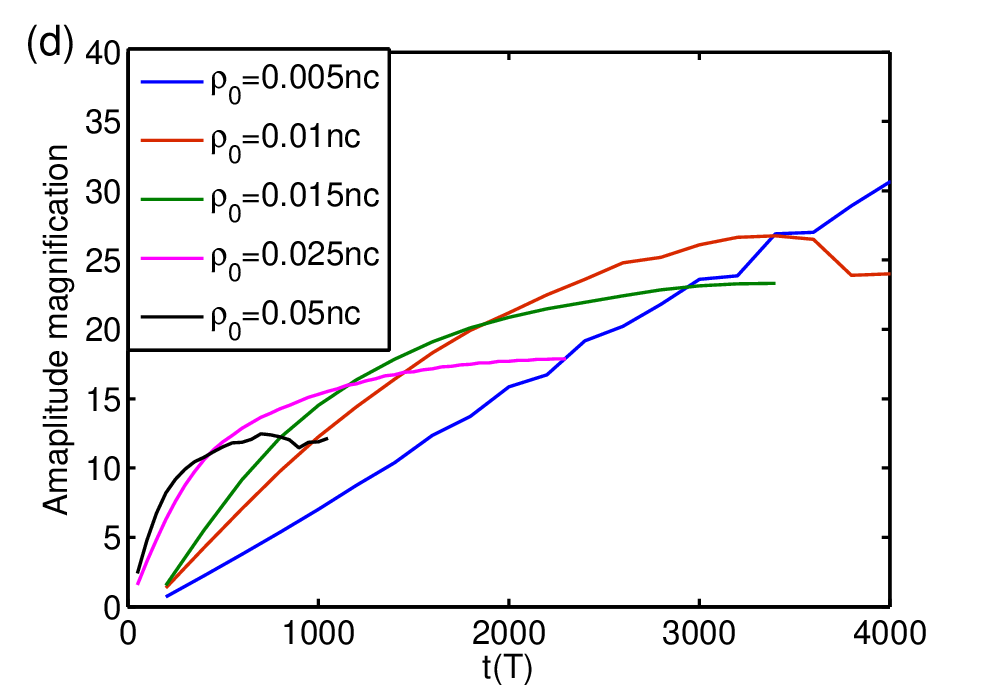}\\
\caption {Simulation results of pulse compression by fast-extending plasma gratings with various pump amplitudes($a_0$), phonon amplitudes($\rho$) and gas densities($\rho_0$) (a)Amplitude magnifications with different pump intensities. (b) Compression radios with different pump intensities. (c) Amplitude magnifications with different phonon amplitudes $\rho$ for $\rho_0=0.01~n_c$, $a_0=0.006$, $\Lambda$ is adjusted according to the maximum AM. (d) Amplitude magnifications with different gas densities $\rho_0$ for $\rho/\rho_0=0.5$, $a_0=0.006$.
\label{fig:optimation}}
\end{figure}

For a higher compressed intensity, the pump amplitude is increased from 0.002-0.007. To compare the compression results , we define amplitude magnification(AM) as the maximum $|b|/a_0$ and compression ratio(CR) as FWHM $t_{Pump}/t_{RP}$. In the beginning stage of compression, AM grows linearly with $t$, which satisfies well to Eq.\ref{SBSM}. As the interaction time becomes longer, it gradually departs from Eq.\ref{SBSM} and then tends to saturation, as shown in Fig.\ref{fig:optimation}(a). The saturation appears earlier with a high pump intensity, probably due to the modulation instability and GVD. As well as AM, the CR also slightly degraded as the pump intensity increases, as shown in Fig.\ref{fig:optimation}(b). However, excellent results are still obtained in a wide range of pump intensity, with optimal AM and CR are above 25 and 1000, respectively. The highest amplitude of 0.17 is obtained by the pump amplitude of 0.007 at $t=3200~T$, corresponding to an intensity of $4\times10^{16}~\rm{W/cm^2}$, which is close to the result obtained by BRA\cite{Mourou12,clark02} at pump wavelength $\sim$1 $\mu$m, but with a much lower pump intensity or interaction length.

The robustness of compression is demonstrated with various $\rho$ and $\rho_0$.  With a larger $\rho$ , the AM grows more quickly, implying the compression radio can be enhanced. However, affected by plasma instabilities, the compression tends to saturation earlier, so the maximum AM  with a longer interaction time does not significantly impacted, as shown in Fig\ref{fig:optimation}(c). Besides, a higher $\rho_0$ may be necessary for a sufficient $\rho$. As $\rho_0$ goes up, the compression can be further accelerated and the reflecting bandwidth for the pump is broadened. However, the modulation instabilities and GVD becomes more serious. The maximum AM and saturated time descends as $\rho_0$ increases, as shown in Fig.\ref{fig:optimation}(d). In the experiment, optimal $\rho$ and $\rho_0$ can be chosen according to the pump duration and bandwidth.

\begin{figure}[htpb]
\centering
\includegraphics[width=1.6 in]{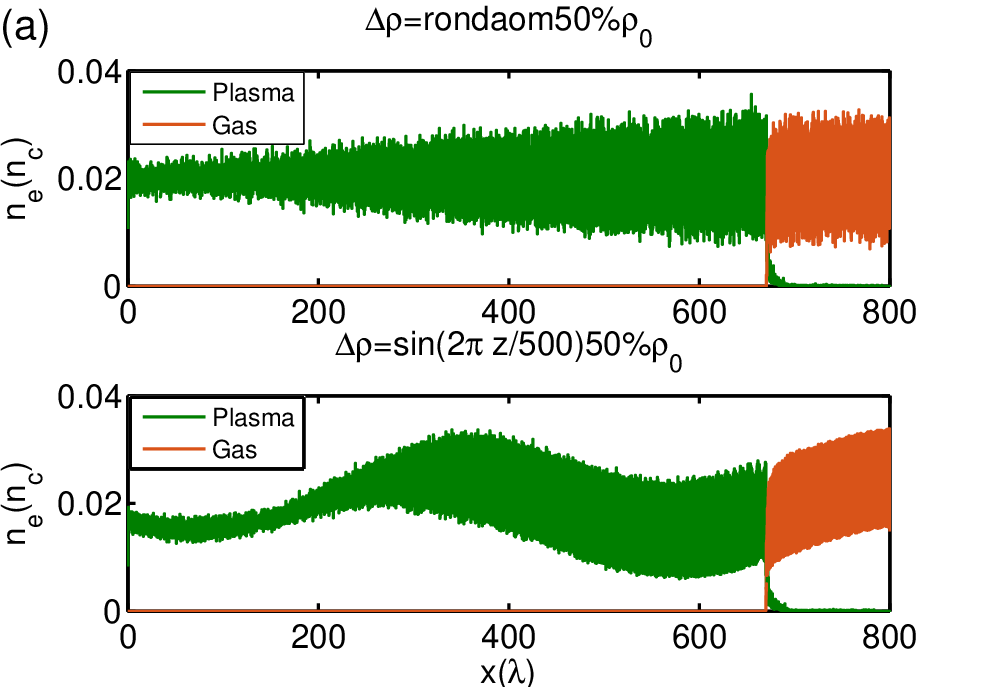}
\includegraphics[width=1.6 in]{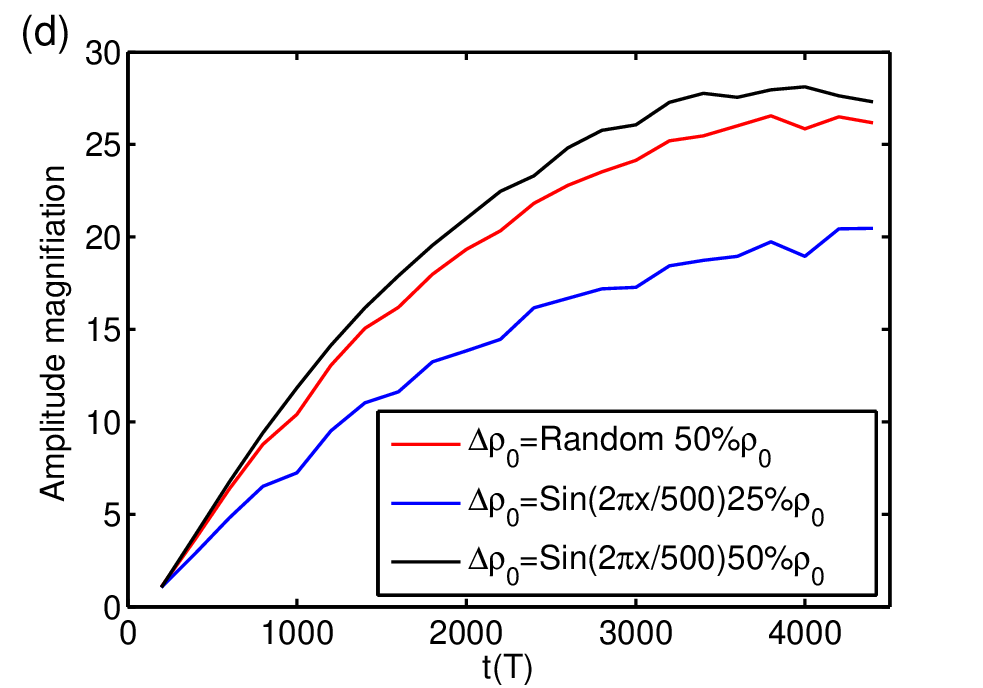}
\caption {PIC Simulation results of FEPGC with inhomogeneous plasmas. (a) Profiles of plasma grating with random and periodic density fluctuations.(b)AM with inhomogeneous plasmas.
\label{fig:densityflu}}
\end{figure}

As well as the Brag gratings in other mediums, the reflectivity is very sensitive to $\rho_0$, so AM drops sharply as $\rho_0$ departs from the Brag condition. The maximum AM decreases from 25 to 5 as $\rho_0$ is reduced by 10\%. However, despite of a fluid medium, it is not difficult to obtain a precise average density $\rho_0$ in plasma. Instead, an unavoidable issue is the density fluctuations. The compression shows a strong robustness against to density fluctuations. For an inhomogeneous density along the laser propagation, the plasma grating is still effective as long as it contains sections satisfying the Brag condition. Therefore, the compression is little influenced by high-frequency density fluctuations, as shown Fig.\ref{fig:densityflu}(b) with a random density flu of $50\%$ $\rho_0$. For a low-frequency fluctuation, the compression divided into 'fast' and 'slow' seasons due to the detuning period is prolonged but broken down. The AM keeps well with a density fluctuations of $50\%\rho_0$ and period of 500$\mu m$, as shown in Fig.\ref{fig:densityflu}(b).

The scheme is easily realized in the experiment. For example, a hypersound wave can be firstly generated by the 1064 nm nanosecond laser pulses in Hydrogen with density of $0.01~n_c$, corresponding to the plasma grating satisfying the Brag condition of an 1053-nm  pump. Then, an 800-nm, 30-fs Ti:Sapphire CPA laser can be used for the ionization pulse, in order to created the FEPG. Finally, laser compression is obtained by using the FEPG to dynamically reflect the Pump. Besides, the generation of FEPG is not limited by using hypersound waves, other paths such a gas jet with periodic structures may also available. Also, the pump wavelength is not limited to around 1 $\mu$m but can extend from extreme ultraviolet(EUV) to microwave by adjusting the acoustic period of the phonon.

In summary, a novel way of laser compression via FEPGC is proposed, showing several excellent features as follows: First, the linear stage is absent as the plasma grating created by ionizing the phonon owns a large amplitude from birth. Consequently, a very high efficiency and short pulse can be obtained. Second, without a growth process, the plasma grating has a strong robustness against plasma instabilities. Third, it is not necessary to prepare a seed as long as the grating and the pump satisfy the Brag condition. In addition, the generation of plasma grating does not depends on the pump, indicating any pump intensity below the ionization threshold works well.

The simulation result of a fully kinetic PIC code Opic displays that a 10-30 ps pump with wavelength of $1~\mu m$ is compressed to a few optical cycles(7-9 fs), having a transfer efficiency near to $80\%$. Supposing a 20-kJ, 20-ps, 1053-nm pump laser is produced by the Nd:glass CPA technique\cite{Mourou12}, it would obtain a peak power around 2 EW($1~\rm{EW}=10^{18}$ W) after the compression. Moreover, as the pulse is compressed to a few optical cycles, it can provide an extremely high-bright source for the generation of attosecond laser pulses.

This work was partly supported by National Key Program for S$\&$T Research and Development (Grant No. 2018YFA0404804), the National Natural Science Foundation of China(Grant No. 11875240) and the Science and Technology on Plasma Physics Laboratory Fund (Grant No. 6142A0403010417).

\bibliography{ref}

\end{document}